\documentclass[10pt]{elsart}
\usepackage{graphicx}

\newcommand{\ket}[1]{|#1\rangle}

\begin{document}
\begin{frontmatter}
\title{Theory of coherent transport by an ultra-cold atomic Fermi gas through linear arrays of potential wells}
\author{M.R.~Bakhtiari},
\author{P.~Vignolo}, 
\author{M.P.~Tosi\corauthref{cor1}}
\corauth[cor1]{Corresponding author, Tel.:+39 050 509275; fax: +39 050 563513. \textit{E-mail address}: {\tt tosim@sns.it}}
\address{NEST-INFM-CNR and Classe di Scienze, Scuola Normale
Superiore, I-56126 Pisa, Italy}
\maketitle


\begin{abstract}
Growing interest is being given to transport of ultra-cold atomic gases through optical lattices 
generated by the interference of laser beams. In this connection we evaluate the phase-coherent transport of a spin-polarized gas of fermionic atoms along linear structures made from potential wells set in four alternative types of sequence. These are periodic chains of either identical wells or pairs of different wells, and chains of pairs of wells arranged in either a Fibonacci quasi-periodic sequence or a random sequence. The transmission coefficient of fermionic matter is evaluated in a T-matrix scattering approach by describing each array through a tight-binding Hamiltonian and by reducing it to an effective dimer by means of a decimation/renormalization method. The results are discussed in comparison with those pertaining to transport by Fermi-surface electrons coupled to an outgoing lead and by an atomic Bose-Einstein condensate. Main attention is given to (i) Bloch oscillations and their mapping into alternating-current flow through a Josephson junction; (ii) interference patterns that arise on period doubling and their analogy with beam splitting in optical interferometry; (iii) localization by quasi-periodic disorder inside a Fibonacci-ordered structure of double wells; and (iv) Anderson localization in a random structure of double wells. 
\end{abstract}

\begin{keyword}
Coherent one-dimensional transport; Quantum gases in optical lattices; Matter-wave optics; Renormalization methods; Fibonacci lattice; Anderson localization.
\PACS{03.75.-b, 05.60.Gg}
\end{keyword}

\end{frontmatter}

\section{Introduction}

Electronic transport in linear arrays of scatterers under an applied electric field has been extensively 
studied on grounds of both intrinsic fundamental interest and relevance to applications. A basic viewpoint was  put forward by Landauer~\cite{1}, who showed that the dc conductance is proportional to the phase-coherent transmittivity. From this viewpoint he proceeded to demonstrate that the resistance of a disordered one-dimensional (1D) lattice grows exponentially with the number of scatterers, as a result of wave-function localization. Within a tight-binding framework the scatterers in the array can in fact be taken to vary according to a chosen law (such as periodic, quasi-periodic, or random) and the density of states and the transmission coefficient can be calculated by Green's function methods (see for instance Economu \cite{2}). A simplifying assumption treats each scatterer as a potential well, thus omitting a detailed account of its internal structure but allowing analytical results to be obtained for linear arrays  (see for instance \cite{3,4,5}).

Advances in the preparation and manipulation of ultra-cold atomic gases have very significantly broadened the scope of the study of phase-coherent quantum transport through arrays of potential wells (for a recent review see \cite{6}). A standing-wave created by the interference of two counter-propagating laser beams, which are detuned away for an atomic absorption line, can be superposed onto an elongated magnetic trap to realise an almost ideal linear periodic array of potential wells for an atomic gas. Higher-dimensional arrays have been created by using two or three pairs of lasers. Experiments on Bose gases in such "optical lattices" have led to the observation of Bloch oscillations under the force of gravity both in ultra-cold atomic gases \cite{7} and in a Bose-Einstein condensate \cite{8}, Landauer-Zener tunneling \cite{9}, Josephson-like oscillations \cite{10,11}, a superfluid to Mott-insulator transition \cite{12}, and the 1D band structure \cite{13}. Collective Bloch oscillations lasting for very many periods have been observed in a spin-polarized Fermi gas inside an optical lattice, opening a novel route to high-precision interferometry and to the measurement of forces with microscopic spatial resolution \cite{14}. The ideal conditions of extremely low temperature and high purity in which experiments on atomic gases can be carried out offers a unique opportunity for the investigation and testing of transport theory at unprecedented levels of accuracy and depth.

A Bose-Einstein condensate that has been adiabatically loaded into an array of potential wells keeps its long-range phase coherence, provided that the barriers between neighboring wells are not too high and still allow tunnel between them \cite{15,16}. In this situation a weak external force accelerates the condensate through the band states of the array as if it were a single quasi-particle \cite{17}. In previous work we have treated coherent transport of an atomic condensate through a variety of linear arrays, using a decimation/renormalization method in a tight-binding approach to reduce each array to an effective dimer \cite{18,19,20}. The arrays that we considered were of four basic types, \textit{i.e.} periodic chains of either identical wells or pairs of different wells and chains of pairs of wells arranged in either a Fibonacci quasi-periodic sequence or a random sequence. In each case we also proposed a specific set-up of optical lasers that through beam interference could create the array in the laboratory. 

In this work we extend the theory to a spin-polarized Fermi gas, this expression being a standard short-hand notation to indicate that the fermionic atoms occupy a single Zeeman sub-level in a magnetic trap. Spin-polarized fermions in an ultra-cold gas can to a very good approximation be regarded as noninteracting, since the antisymmetry of the many-body wave function under exchange that is embodied in the Pauli principle inhibits close encounters between them and this suppresses \textit{s}-wave collisions (for a full discussion see for instance \cite{6}). 
The absence of interactions makes the spin-polarized Fermi gas a good
candidate to observe Anderson localization of matter waves.
In the ground state the fermions occupy the single-particle levels of the confining potential up to a maximum fixed by their number, and in particular in a 1D single-well periodic lattice at half filling they fill the bottom half of the lowest energy band. This situation should be contrasted with that of a Bose-Einstein condensate, where the gas in its ground state is condensed in the lowest (zero quasi-momentum) band state and, as already noted, is driven by a constant external force through the band states as if it were a single quasi-particle. Coherent motion of a spin-polarized Fermi gas under an imposed bias can still occur if each fermion is in turn accelerated into successively higher band states. The whole distribution is thereby coherently accelerated through the band structure in the absence of dissipative scattering. 

The contents of the paper are briefly as follows. In Section \ref{sec-2} we present the model and discuss the essential aspects of the density of states and of the calculation of the transmission coefficient. Numerical results are reported in Section \ref{sec-3} for a spin-polarized gas of fermionic $^{40}$K atoms, this being the system for which long-lasting Bloch oscillations in a single-well optical lattice have been observed by Roati \textit{et al.} \cite{14}. Our results are discussed in comparison with those for coherent transport by Fermi-level electrons \cite{5} and by a Bose-Einstein condensate \cite{18,19,20}. The main points at issue are (i) the Bloch oscillations of the gas and their mapping into current flow across a Josephson junction; (ii) the interference patterns between Bragg scattered inter-subband tunneling fragments that arise on period doubling; (iii) the localization of fermionic matter induced by the introduction of quasi-periodicity; and (iv) Anderson localization induced by randomness. Finally, Section \ref{sec-4} summarizes our conclusions.

\section{The tight binding model and the array transmittivity}\label{sec-2}
We consider a 1D sequence of $N$ equally spaced potential wells to be occupied by $N_f$ spin-polarized fermions. In a tight binding model the Hamiltonian is
\begin{equation}\label{eq:H}
H=\sum_{i=1}^N \{ e_i\,
|\,i\rangle\,\langle i\,|\,+(t_{i,i+1}\,|\,i\rangle\,\langle
i+1|+t_{i+1,i}|\,i+1\rangle\,\langle i\,|\,) \}
\end{equation}
where the site energy $e_i$ corresponds to the lowest energy level in the $i$-th well and $t_{i,i\pm 1}$ are the hopping energies between adjacent wells. The case of practical interest for an atomic gas in a 1D optical lattice is that in which a few handred sites are occupied. The calculation reported in Section \ref{sec-3} below will refer to $^{40}$K atoms at half filling, with $N_f=100$ and $N=200$.

The array is described by an external potential $U(z)$ constructed from the periodic function $\sin^2(\pi  z/d)$, $d$ being the distance between adjacent wells, and a bimodal distribution of well depths ($U_1$ and $U_2$, say). The potential created on an atomic gas by a standing optical wave generated from two counter-propagating laser beams correspond to $U_2=U_1$ and $d$ equal to one-half the laser-light wavelength. In addition to this case, in which $U(z)$ describes a period sequence of identical wells, we also treat cases in which the $U_i$'s are chosen so that the well depth (i) regularly alternate along the array, or (ii) form a Fibonacci sequence, or (iii) form a sequence produced by a random-number generator.

The global density of states of each array is obtained from the Green's function of the model by means of the Kirkman-Pendry relation \cite{21}, expressing coherence between the first and the last site of the chain. Illustrative examples have been given in our previous work \cite{5}. Of course, period doubling breaks the lowest energy band of the chain into two sub-bands separated by an energy gap, whereas the introduction of quasi-periodic or random disorder induces spectral fragmentation and a multiplicity of pseudo-gaps.

\begin{figure}
\includegraphics[width=1.1\linewidth]{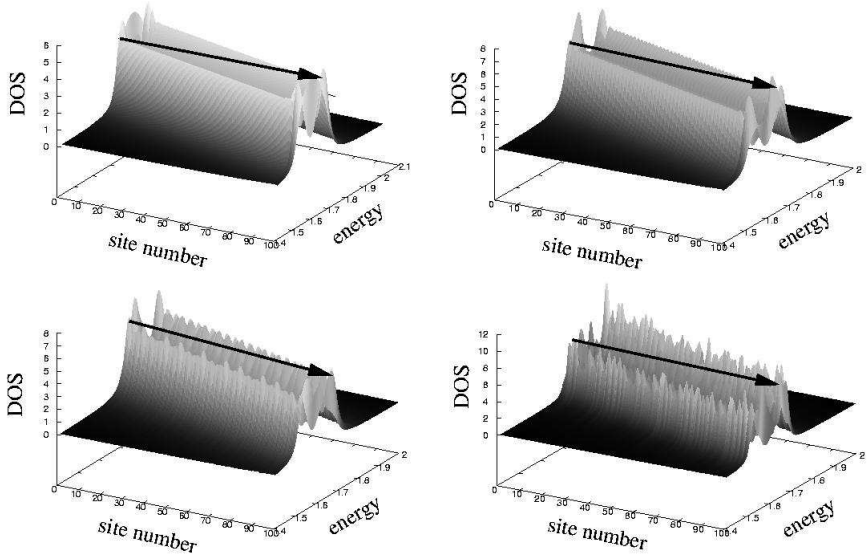}
\caption{Local density of states as a function of position and energy for a single-well, a periodic double-well, a Fibonacci, and a random array (from top left to bottom right). The arrow indicates the trajectory of an atom starting at a given initial energy.}
\label{fig-1}
\end{figure}

An applied constant force causes a linear tilt of the spectrum and the atoms travel along the chain and at the same time are accelerated through the energy spectrum \cite{22}. A schematic representation of the trajectory of an atom moving in space and energy can be obtained from the local (site projected) density of states $D_i(E)$ (LDOS) and is shown by the arrows in Fig. \ref{fig-1} for a chosen value of the initial energy. The LDOS is given by the expression
\begin{equation}
D_i(E)=-\frac{1}{\pi}
\,{\rm Im}\,\langle i|(E-H)^{-1}|i\rangle\,\equiv -\frac{1}{\pi}
\,{\rm Im}\,G_{i,i}(E)
\end{equation}
where $G_{i,i}(E)$ is the site-projected Green's function at energy $E+i\eta$
in the limit of vanishing positive $\eta$. This matrix element can be evaluated as $G_{i,i}(E)=[E-\tilde{\varepsilon}_i(E)]^{-1}$ by reducing the array to an effective site with energy $\tilde{\varepsilon}_i(E)$ through decimation of all $j$-th sites with $j\neq i$, the renormalized Hamiltonian being a $c$-number expressed in term of a continued fraction \cite{23,24,25,26,27}. The LDOS in Fig. \ref{fig-1} shows that at a given energy the population of a subset of sites is favored in both the double-well and the disordered arrays, these sites being rather irregularly distributed along the array in the latter case \cite{19}.

The same renormalization procedure is used in the calculation of the transmittivity to reduce each array to an effective dimer containing just the first and the last site with renormalized site and hopping energies \cite{23,24,25,26,27}. The Hamiltonian $\tilde{H}(E)$ of the dimer depends on the energy $E$ and is expressed as a $2\times 2$ matrix,
\begin{equation}
\tilde{H}(E)=
 \left(
\begin{array}{ccc}
\tilde{\varepsilon}_1^{(N-2)}(E) & & \tilde{t}_{1,N}(E) \\
 \tilde{t}_{N,1}(E)& &\tilde{\varepsilon}_N^{(N-2)}(E)\\
\end{array}
\right).
\label{H}
\end{equation}
Here, the effective site and hopping energies are determined by the recursive relations
\begin{eqnarray}
\tilde{\varepsilon}_1^{\,(j)}(E)&=&\tilde{\varepsilon}_1^{\,(j-1)}(E)+
\tilde{t}_{1,j+1}(E)\,\frac{1}{E-\tilde{\varepsilon}_{j+1}^{\,(j-1)}(E)}
\,\,\,t_{j+1,j+2}\,,\label{eq:rec1}\\
\tilde{\varepsilon}_{j+2}^{\,(j)}(E)&=&e_{j+2}+t_{j+2,j+1}
\,\frac{1}{E-\tilde{\varepsilon}_{j+1}^{\,(j-1)}(E)}\,\,
\tilde t_{j+1,1}(E)\,,\\
\tilde{t}_{1,j+2}(E)&=&\tilde{t}_{1,j+1}(E)\,
\frac{1}{E-\tilde{\varepsilon}_{j+1}^{\,(j-1)}(E)}\,\,\,t_{j+1,j+2}\,,
\label{eq:rec3}
\end{eqnarray}
and $\tilde t_{j+1,1}=\tilde t_{1,j+1}$ for $j\ge 1$. The initial values
are given by the original Hamiltonian parameters, 
$\tilde{\varepsilon}_i^{\,(0)}(E)=e_i$ and $\tilde t_{1,2}(E)=t_{1,2}$.

\subsection{The transmittivity} 
We can now proceed to calculate the transmittivity of the array for a gas driven by a constant force $F$. For this purpose we calculate the scattering matrix in the presence of the bias, in a configuration in which the array is connected to incoming and outgoing leads \cite{18}. The leads mimic a coupling to the continuum by injecting and extracting a steady-state particle current. In describing an experiment such as that  carried out by Anderson and Kasevich \cite{8} on a Bose-Einstein condensate, the incoming lead would correspond to a continuous replenishing of the condensate in the array and the outgoing lead to a detecting system counting the atoms that leave the array in coherent drops separated in time by the Bloch-oscillation period under gravity. In an experiment such as that carried out by Roati \textit{et al}. \cite{14} on a Fermi gas, instead, the leads should be blocked after loading the gas in the array.

\begin{figure}
\includegraphics[width=0.9\linewidth]{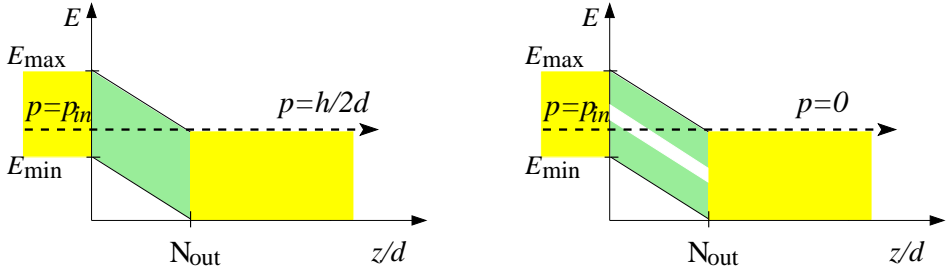}
\caption{Band tilting in space for a single-well array (left) and
a periodic double-well array (right), due to a constant force $F$. Coupling
to an incoming and an outgoing lead is also shown for an atom with initial momentum $p_{\mathrm{in}}$ and exiting after $N_{\mathrm{out}}$ hops at the Brillouin zone edge (left) or at the Brillouin zone center (right). The position $z$ is in units of $d$.}
\label{fig-2}
\end{figure}

With specific reference to a spin-polarized atomic Fermi gas, all fermions with initial momentum $p_{\mathrm{in}}$
in the range $p_{\mathrm{in}}\in [-p_F,p_F]$, with $p_F=(2mE_F)^{1/2}$ being the Fermi momentum, move coherently under the applied force in the absence of dissipative scattering processes. We define $N_{\rm out}=(E_{\mathrm{max}}-E_{in})/(Fd)$ (with $N_{\rm out}\le N$) as the number of hops
after which an atom having initial energy $E_{in}$ and
momentum $p_{\mathrm{in}}$ reaches the highest  point in the dispersion curve at energy $E_{\mathrm{max}}$  and can either leave the array through the outgoing lead or continue its trajectory through the band states (see Fig. \ref{fig-2} for an illustration referring to the periodic single-well and double-well arrays). The leads are defined by two additional terms in the Hamiltonian,
\begin{equation}
H_{\textrm{\begin{scriptsize}L,in\end{scriptsize}}}=\sum_{n=-\infty}^0\,\{E_0^{\rm in}|n\rangle\langle n|+
t_0(|n\rangle\langle n+1|+\textrm{c.c.})\}
\label{lead_l}
\end{equation}
and
\begin{equation}
H_{\textrm{\begin{scriptsize}L,out\end{scriptsize}}}=\sum_{n=N+1}^\infty\,\{E_0^{\rm out}|n\rangle\langle n|+
t_0(|n-1\rangle\langle n|+\textrm{c.c.})\}\,.
\label{lead_r}
\end{equation}
We take the hopping energy $t_0$ equal to one fourth of the spectral band width and shift the centers of the leads by $F d N_{\mathrm{out}}$ in order to optimize their coupling to the array.

The outgoing wave function $|\varphi_{\rm out}^{\tilde\kappa}\rangle$
is an eigenstate of the outgoing lead and
is related to the incoming wave function $|\varphi_{\rm in}^\kappa\rangle$, 
which is an eigenstate of 
the incoming lead, by 
\begin{equation}
|\varphi_{\rm out}^{\tilde\kappa}\rangle=
(1+G^0T)|\varphi_{\rm in}^\kappa\rangle\,.
\label{omegapiu}
\end{equation}
The wave vectors $\kappa$ and $\tilde{\kappa}$ of the two wave functions are uniquely determined  by the relations
\begin{equation}
\kappa,\tilde{\kappa}=\frac{1}{d}
\arccos\left(\frac{E-E_0^{\rm in,out}}{2t_0}\right).
\end{equation}
Further, in Eq. ({\ref{omegapiu}) $G^0=(E-H_0)^{-1}$ is 
the Green's function 
of the disconnected leads, described by the Hamiltonian
$H_0=H_{\textrm{\begin{scriptsize}L,in\end{scriptsize}}}+H_{\textrm{\begin{scriptsize}L,out\end{scriptsize}}}$, 
and the scattering matrix 
$T(E)=H_\textrm{\begin{scriptsize}I\end{scriptsize}}(1-G^0H_\textrm{\begin{scriptsize}I\end{scriptsize}})^{-1}$ 
referred to the perturbation Hamiltonian 
\begin{equation}
H_{\rm I}=\tilde H|_{N=N_{\rm out}}-(E_0^{\rm in}\,|1\rangle\langle 1|+
E_0^{\rm out}\,|N_{\rm out}\rangle\langle N_{\rm out}|),
\end{equation} 
with $\tilde{H}|_{N=N_{\mathrm{out}}}$ being given by Eq. ({\ref{H}) for the case $N=N_{\mathrm{out}}$.

Since the wave functions $|\varphi_{\rm in}^{\tilde\kappa}\rangle$ and $|\varphi_{\rm out}^{\tilde\kappa}\rangle$ are defined in disconnected spaces, the projection of $|\varphi_{\rm out}^{\tilde\kappa}\rangle$ onto the localized function $\ket{n}$ can be written as
\begin{eqnarray}
\langle n\,|\,\varphi_{\rm out}^{\tilde\kappa}\rangle&=&
\sum_{j\le1,l}G^0_{n,l}T_{l,j}\langle j|\varphi_{\rm in}^\kappa\rangle \nonumber\\
&=&\sqrt{2}G^0_{n,N_{\rm out}}T_{N_{\rm out},1}
\sin(\kappa d)
u_{\kappa}(d)
\label{anna2}
\end{eqnarray}
where we have set $\langle 1\,|\varphi_{\rm in}^\kappa\rangle=u_{\kappa}(d)
\big(e^{i\kappa d}-
e^{-i\kappa d}\big)/(i\sqrt{2})$ with $u_{\kappa}(d)$ being the Wannier function in the potential
$U(z)$.
The Green's function element $G^0_{n,N_{\rm out}}$ in Eq.(\ref{anna2})
determines the coherence
between site $n$ and site $N_{\rm out}$ on the chain
for the outgoing lead and can be written as
\begin{equation}
G^0_{n,N_{\rm out}}=\frac{t_0^{n-N_{\rm out}}}
{|t_0^{n+1-N_{\rm out}}|}\exp[\,{i\,\tilde{\kappa}(n+1-N_{\rm out})\,d}\,]\,.
\end{equation}
In the out-of-equilibrium picture,
the velocities $v_{\rm in}$ and $v_{\rm out}$ of the incoming and
outgoing wave functions enter the definition of 
the transmission coefficient ${\mathcal T}$ 
as 
\begin{equation}
{\mathcal T}=|\tau|^2
\frac{v_{\rm out}}{v_{\rm in}}
=\frac{\lim_{n\rightarrow+\infty}
\langle n|\varphi_{\rm out}^{\tilde\kappa}\rangle
\langle \varphi_{\rm out}^{\tilde\kappa}|n\rangle v_{\rm out}}
{\lim_{m\rightarrow-\infty}\langle m|\varphi_{\rm in}^\kappa\rangle
\langle \varphi_{\rm in}^\kappa|m\rangle v_{\rm in}}.
\end{equation}
Finally we obtain
\begin{equation}
{\mathcal T}=4\frac{|T_{1,N_{\rm out}}|^2}{t_0^2}\,
\sin{\left(\kappa d\right)}
\sin{\left(\tilde{\kappa}d\right)}.
\label{elei}
\end{equation}
We remark that  Eq. (\ref{elei}) reduces to the transmission coefficient evaluated in \cite{5} for
 Fermi-level electrons when we set $E_0^{\rm out}=E_0^{\rm in}$.

\section{Numerical results} \label{sec-3}
Our task now is to determine the parameters $e_i$ and $t_{i,i+1}$ entering the Hamiltonian in Eq. (\ref{eq:H}). We proceed for this purpose to a 1D reducing of the system by introducing the transverse width $\sigma_\perp$ of the atomic cloud in an elongated cigar-shaped harmonic trap and the 1D wave function in the $i$-th well. In a tight binding scheme the Wannier function $\Psi_i(z)$ in the potential $U(z)$ can be written, according to the work of Slater \cite{28}, as a Gaussian function having a width $\sigma_{z_i}$ determined by the harmonic approximation to the well. That is,
\begin{equation}
\Psi_i(z)=(\pi^{1/4}\sigma_{{z_i}})^{-1}
\exp[-(z-z_i)^2/(2\sigma_{z_i}^2)]
\end{equation}
where $\sigma_{z_i}=(\hbar/m\omega_i)^{1/2}$ with $\omega_i=(4 U_i E_R)^{1/2}$, $E_R$ being the recoil energy. 
The site energies are then given by
\begin{equation}
e_i=\int dz\,\Psi_i(z)\left[
-\frac{\hbar^2\nabla^2}{2m}+U(z)+maz+C\right]\Psi_i(z).
\label{siteenergy_b}
\end{equation}
Here, $a=F/m$ is the acceleration due to the constant force $F$ applied to the fermions, and $C=\hbar^2/(2m\sigma_{\perp}^2)+\frac{1}{2}m\omega_{\perp}^2\sigma_{\perp}^2$ is a quantity taking into account the reduction to 1D dimensionality, with $\omega_\perp$ being the radial frequency of the trap \cite{29}. 
The hopping energy are similarly given by
\begin{equation}
t_{i,i+1}=\int dz\,\Psi_i(z)
\left[
-\frac{\hbar^2\nabla^2}{2m}+U(z)+C\right]\Psi_{i+1}(z).
\label{hopenergy_b}
\end{equation}

For a numerical illustration we have chosen to adopt typical system parameters used in
experiments on atomic gases, by setting $U_1=3.5E_R$ and $2d=763$\,nm with $(U_2-U_1)/U_1=10^{-2}$ in the double-well arrays.  
Our results are shown in Fig. \ref{fig-3} where we have plotted the transmittivity of the spin-polarized Fermi gas of $^{40}$K atoms through periodic arrays and through a Fibonacci-ordered and a random array of double wells, as a function of the inverse of the applied force and of the energy. In all cases the array is at half filling and the inverse force is described through the quantity $T_B/\tau_t$, where $T_B=h/(2Fd)$ is the Bloch-oscillation period for the fermions in the double-well lattice and $\tau_t$ is a time-scale parameter to be defined immediately below. The energy variable is written as the ratio $(E-E_{\mathrm{min}})/(E_F-E_{\mathrm{min}})$.

\begin{figure}
\begin{center}
\includegraphics[width=0.9\linewidth]{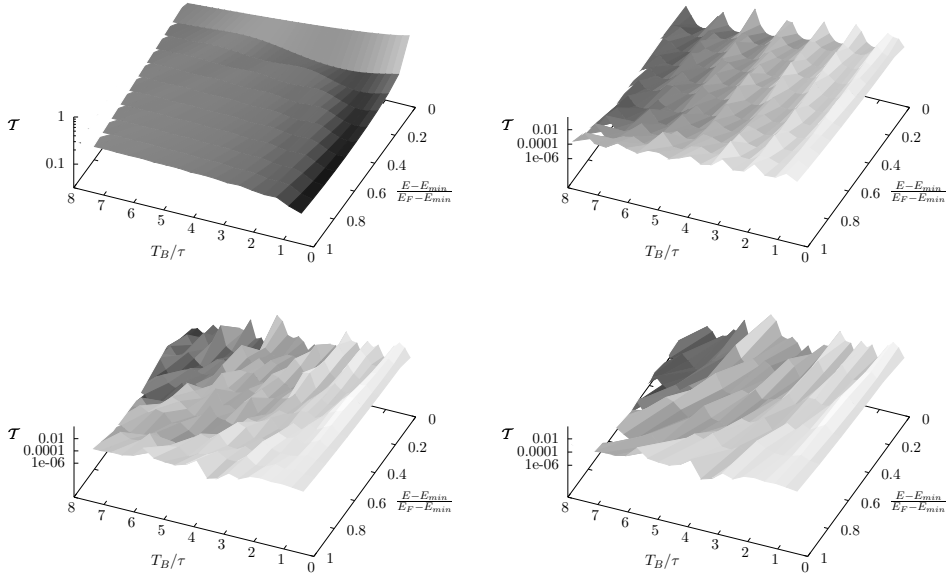}
\caption{Transmittivity $\mathcal{T}$ (in log scale) of a spin-polarized Fermi gas through a single-well lattice (top-left), a periodic double-well array (top-right), a Fibonacci-ordered double-well array (bottom-left), and a random double-well array (bottom-right). The transmittivity at half filling is plotted as a function of $T_B/\tau_{t}$ and of $(E-E_{\mathrm{min}})/(E_\mathrm{F}-E_{\mathrm{min}})$. The time scales $T_B$ and $\tau_t$ in all cases are taken as the Bloch oscillation period and the inter-subband tunneling time in the double-well lattice.
\label{fig-3}}
\end{center}
\end{figure}

It is seen from Fig. \ref{fig-3} that fermions at the Fermi level in the single-well lattice have a better coupling with the outgoing lead at low values of the drive, while those at very low energies have a greater coefficient of transmission at large values of the drive. The behavior of the transmittivity through the double-well lattice is illustrated in the top-right panel of Fig. \ref{fig-3}. We expect in this case destructive interference in the transmission whenever the Bloch period is an integer multiple of the time to tunnel twice across the inter-subband minigap $\Delta E$. The underlying phenomenon is the interference occurring at $p=-h/(4d)$ (the left-hand boundary of the Brillouin zone in the split-band configuration) between the atomic wave packets that are Bragg scattered at $p=h/(4d)$ in the lower sub-band and those that have tunneled into the upper sub-band, traveled through it, and tunneled back into the lower sub-band  at $p=-h/(4d)$. From our numerical data we determine the fermion tunneling time as $\tau_t=6\pi\hbar/\Delta E$. A proportionality of the tunneling time to $\hbar/\Delta E$ is expected from basic quantum mechanics (see for instance \cite{30}), but the specific value that we find for the proportionality constant naturally depends on the model Hamiltonian that we have assumed. 

Turning to the effect of quasi-periodic or random disorder, we first remark that the fermions traveling through the array of Fibonacci-ordered double wells explore an energy spectrum which on average is rather more  akin to the single-well band structure than to the double-well one, but is regularly fragmented by a multiplicity of pseudo-gaps \cite{5}. The quasi-periodic disorder nevertheless induces marked peaks and troughs in the transmission as can be seen in the bottom-left panel of Fig. \ref{fig-3}. The peak positions do not show any regularity and depend on both the applied drive and the fermion energy. Finally, the bottom-right panel in Fig. \ref{fig-3} shows that random disorder in a double-well array at half occupancy induces an interference pattern which is qualitatively similar
to that of a Fibonacci array.

With the aim of making a direct comparison  of our results for transport in a Fermi gas with the transmittivity of a Bose-Einstein condensate calculated in previous work \cite{18,19,20}, we have plotted in Fig. \ref{fig-4} the mean transmittivity of the Fermi gas after averaging over incoming states, namely the quantity 
\begin{equation}
\bar{\mathcal{T}}=[\int_{E_{\mathrm{min}}}^{E_\mathrm{F}} \mathcal{T}(E)D_{\mathrm{in}}(E)\, dE\, ]/[\int_{E_{\mathrm{min}}}^{E_\mathrm{F}} D_{\mathrm{in}}(E) \,dE \,]
\end{equation}
as a function of $T_B/\tau_t$, $D_{\mathrm{in}}(E)$ being the density of states of the incoming lead. The behavior of the averaged fermion transmittivity is qualitatively similar to that found for bosons. In particular, Bloch oscillations with a period $h/(Fd)$ are executed by the whole distribution of fermions once the coupling to the leads is cut. This type of behavior was observed for a condensate in the experiments of Anderson and Kasevich \cite{8} and was interpreted in term of the condensate moving through the band states as a single coherent quasi-particle \cite{15,18}. Our calculations confirm that a similar coherent-transport behavior, as observed in the experiments of Roati \textit{et al}. \cite{14}, is exhibited by a spin-polarized Fermi gas in a single-well lattice, even though the fermions start from a ground-state configuration in which they are distributed over many band states up to the Fermi level instead of being all condensed in the lowest band state at the Brillouin zone center.

\begin{figure}
\begin{center}
 \includegraphics[width=0.47\linewidth]{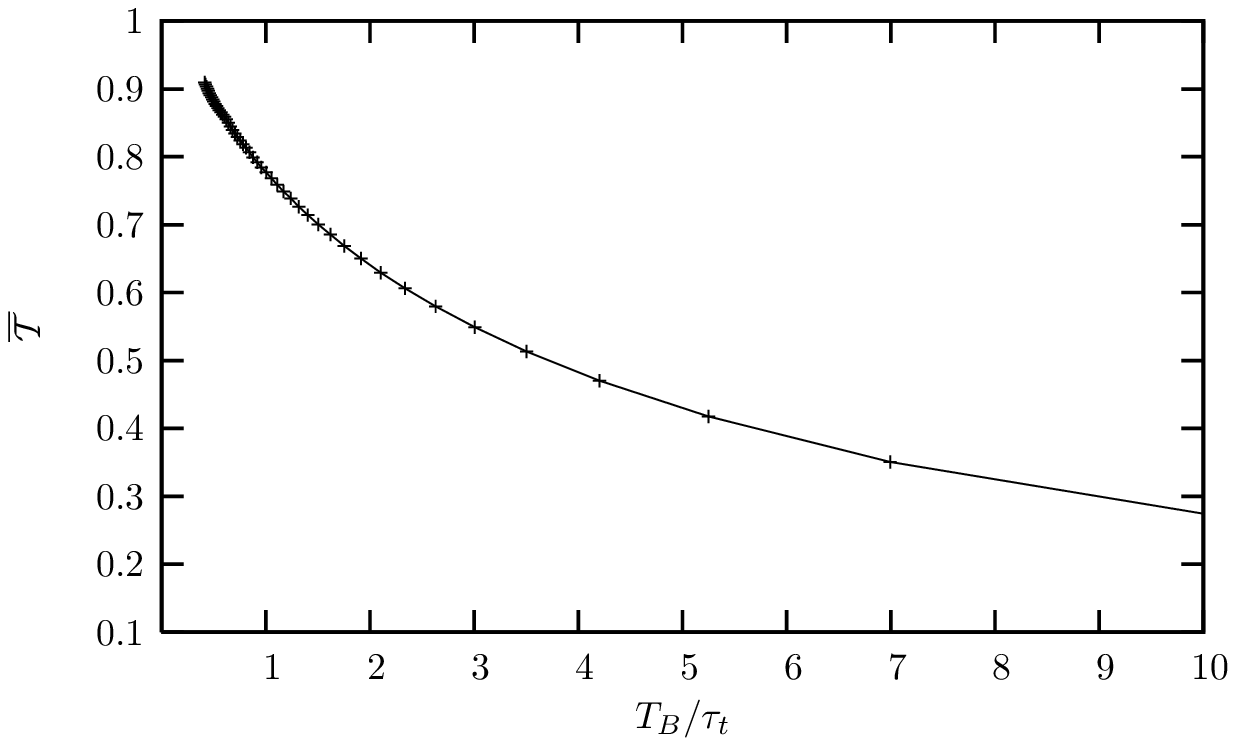}
 \includegraphics[width=0.47\linewidth]{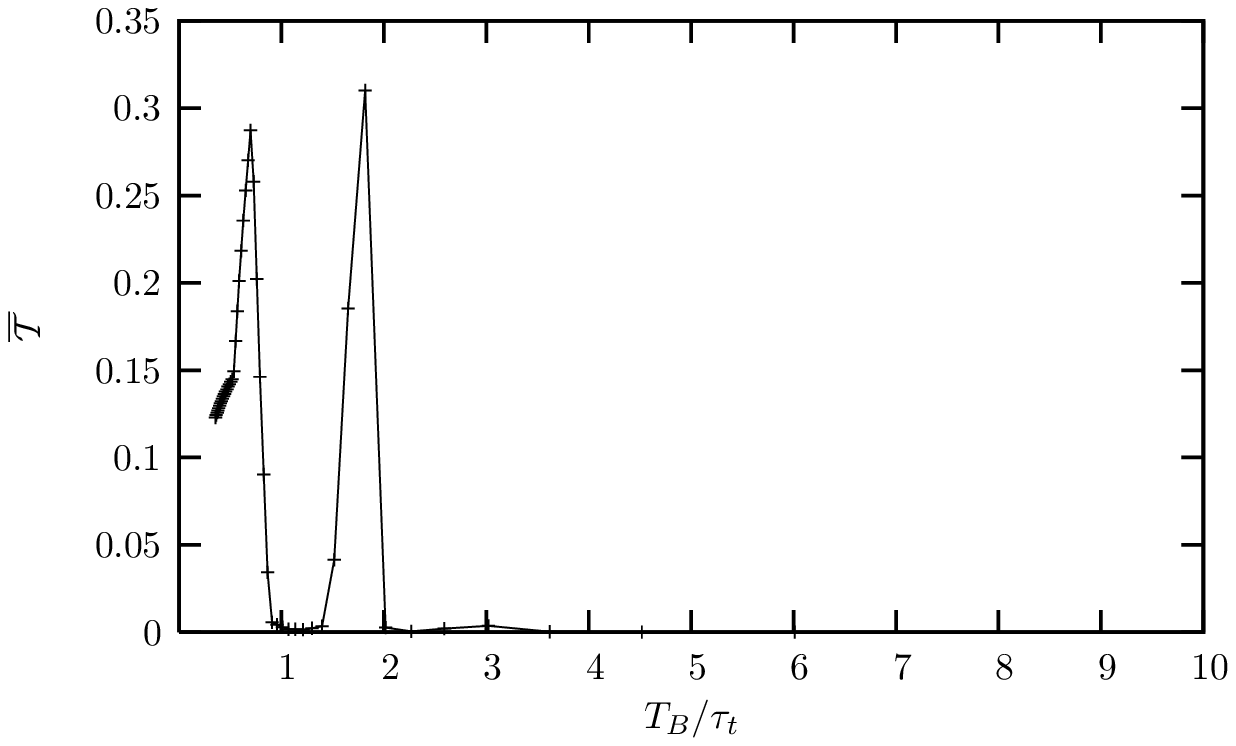}
 \includegraphics[width=0.47\linewidth]{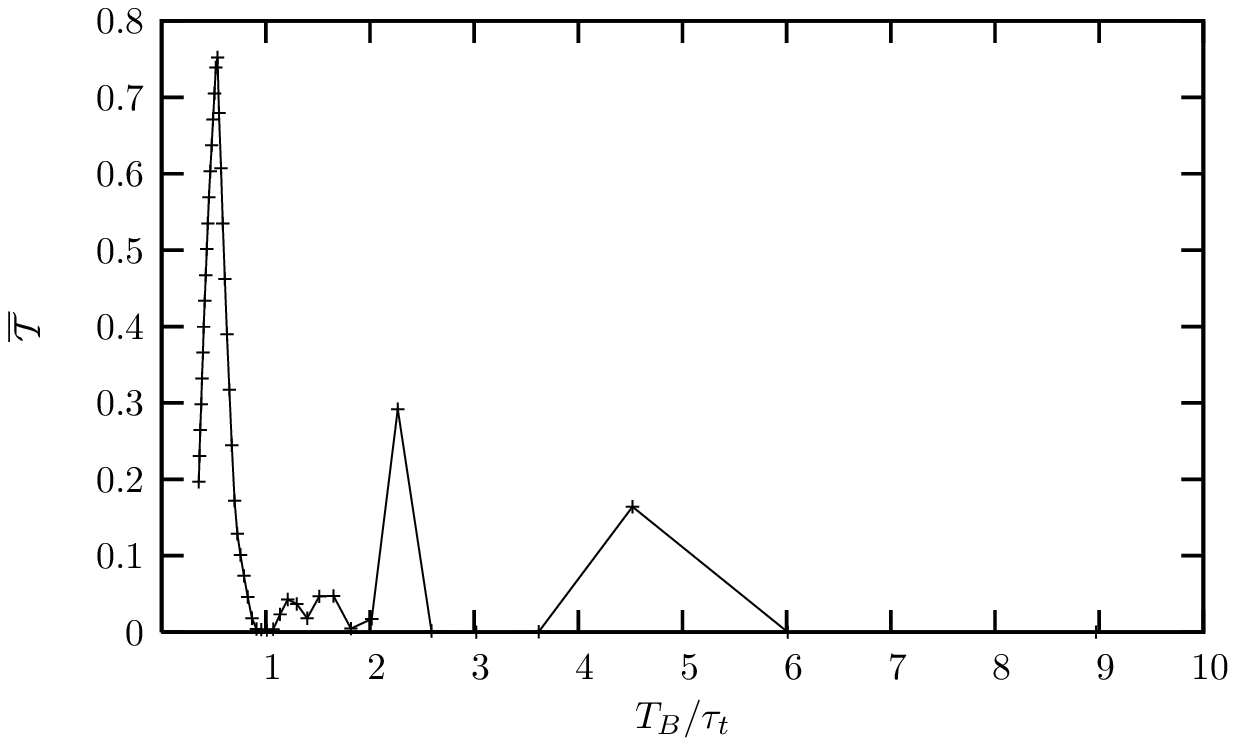}
 \includegraphics[width=0.47\linewidth]{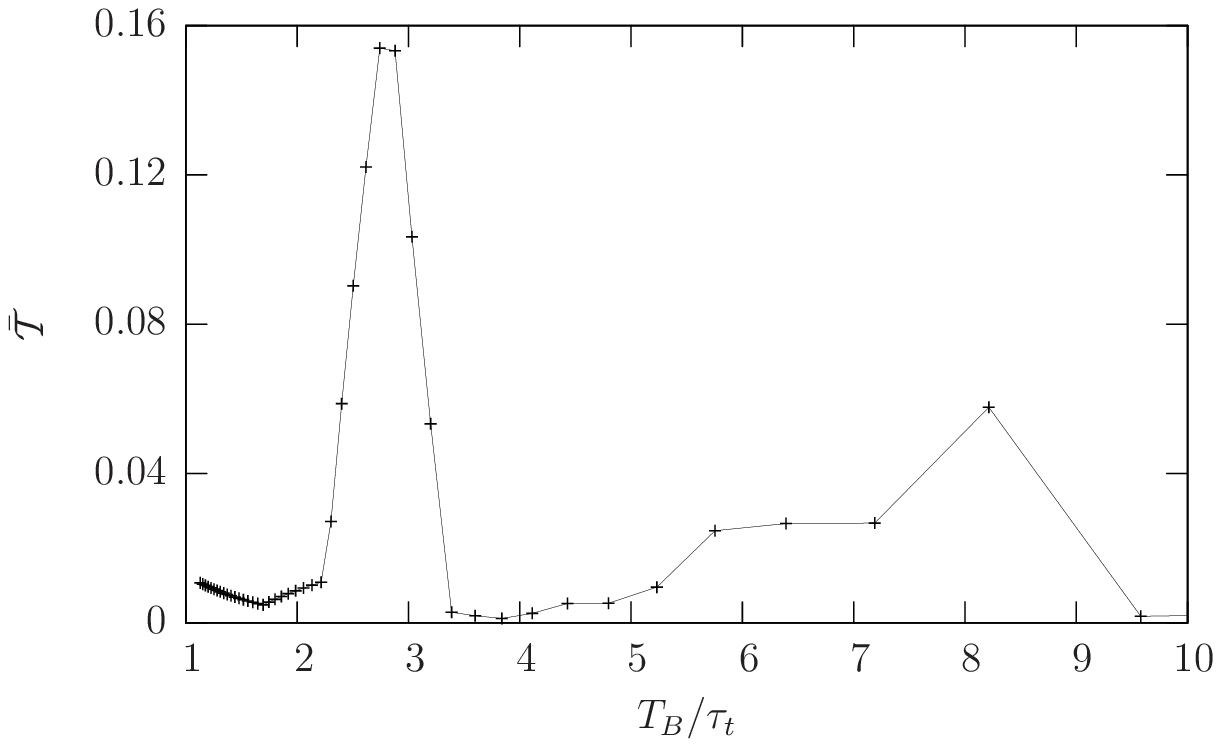}
\caption{Averaged transmittivity $\bar{\mathcal{T}}$ of a spin-polarized Fermi
gas at half filling as a function of $T_B/\tau_{t}$. The various panels refer to a single-well lattice  (top-left), a periodic double-well array (top-right), a Fibonacci-ordered double-well array (bottom-left), and a random double-well array (bottom-right).\label{fig-4}}
\end{center}
\end{figure}

However, it is also seen from Fig. \ref{fig-4} in comparison with previous results for a Bose-Einstein condensate \cite{18,19,20} that inter-subband coherence is sharply reduced in a Fermi gas and that the consequences of quasi-periodic or random disorder in a double-well array are more dramatic. The low values in the transmission coefficient in the presence of disorder essentially correspond to 
Anderson localization of the fermionic matter induced by the large number of pseudo-gaps that have opened up in the density of states. The non-zero values of the transmittivity are due to finite-size effects.

\section{Conclusions and future perspectives} \label{sec-4}
In summary, we have focused attention in this work on some main features of the coherent transport of gaseous ultra-cold fermionic matter through linear arrays of potential wells under a constant external drive. Our present  results reduce in the case of infinitesimal bias to those holding for transport by Fermi-level electrons through arrays of confining sites such as quantum dots, as evaluated in previous work \cite{3,4,5}. The discussion of our results has been framed within a comparison with those for coherent transport of bosonic matter by a Bose-Einstein condensate \cite{15,16,17,18,19,20} with main attention to long-lasting Bloch oscillations in a lattice of confining sites, to matter-wave interference induced by splitting through an inter-subband minigap and having an analogue in optical beam-splitting experiments, and to localization of matter waves by quasi-periodicity and by random disorder. We have used a procedure of site decimation  followed by renormalization, which allows analytic development of the theory for linear arrays and could usefully be extended to transport in networks of confining sites having higher dimensionality.

In comparison with the many experimental studies that have been carried out on electronic transport through arrays or networks of quantum dots, the experimental study of matter transport by quantum atomic or molecular gases through optical arrays is still in its infancy though promising of novel fundamental developments. Both bosons and fermions have been observed to perform Bloch oscillations inside a single-well optical lattice \cite{8,14} and in both cases the experiments give clear evidence for long-range coherence of matter waves. Transport of quantum matter through other types of array remains to be studied experimentally. A most promising development would seem to lie in the direction of one-dimensional transport by two-component Fermi gases consisting of two spin populations, forming a Luttinger liquid for repulsive interspecies interactions and leading to Luther-Emery spin pairing for attractive interactions \cite{31}. Separate control of the two spin populations has been achieved \cite{32,33} and opens the way to the study of exotic superfluid states \cite{34,35,36,37,38}.

\end{document}